\documentclass[aps,pre,superscriptaddress,twocolumn,letter,citeautoscript]{revtex4}
\usepackage{graphicx}
\usepackage{times}

\begin{document}
\title{The infrared spectra of ABC-stacking tri- and tetra-layer graphenes \\
studied by first-principles calculations}

\author{Yuehua Xu}
\author{San-Huang Ke} 
\email[Corresponding author, E-mail: ]{shke@tongji.edu.cn}
\affiliation{Department of Physics, Tongji University, 1239 Siping Road, Shanghai 200092, P. R. of China}

\begin{abstract}
The infrared absorption spectra of ABC-stacking tri- and tetra-layer graphenes
are studied using the density functional theory. It is found that they exhibit
very different characteristic peaks compared with those of AB-stacking ones,
caused by the different stacking sequence and interlayer coupling. The
anisotropy of the spectra with respect to the direction of the light electric
field is significant. The spectra are more sensitive to the stacking number when
the electric field is perpendicular to the graphene plane due to the interlayer
polarization. The high sensitivities make it possible to identify the stacking
sequence and stacking number of samples by comparing theory and experiment.  
\end{abstract}
\maketitle

\section{Introduction}

Single-layer graphene (SG) is a two-dimensional (2D) flat monolayer composed of
carbon atoms arranged in a honeycomb network, which is a basic building block
for all other graphite materials. In the past, it was believed to be unstable in
free-standing state \cite{1} and so described as an 'academic' material. However, SG was
unexpectedly made experimentally several years ago \cite{2}. The unique electronic
properties of SG is mainly due to its very peculiar band structure, with the
$\pi$ and $\pi^{*}$ bands showing linear dispersion around the Fermi level
($E_{\rm F}$) where they
touch with each other at a single point $K$ in the Brillouin zone (BZ) \cite{3,4}.
Because of its great potential for applications in nano-science and technology,
SG has been thoroughly studied both theoretically and experimentally
\cite{5,6,7,8,9,10}. 

Recently, a further progress in this field was the successful fabrication and
use of few-layer graphenes (FGs) which are stacking of a few graphene layers
\cite{11,12,13}. From the viewpoint of application, FGs can be even more useful than SG
since it offers further control of electronic states by interlayer
interactions \cite{14,15,16,17,18} , or by applying an electric field perpendicular to the
molecular plane to open a band gap which is critical for applications in
nanoelectronics \cite{19,20}. Because of the interlayer interaction, theoretically,
there can be dramatic changes in the electronic properties of FGs compared with
those of SG, depending on the stacking sequence and number ($N$). Considering an
arbitrary arrangement for adjacent graphene layers, there will be 2$^{N-2}$ possible
low-energy stacking sequences for FGs of $N$ layers. It is therefore desirable and
important to understand the electronic properties of the different FGs and to be
able to identify accurately their stacking sequence and number. Experimentally,
infrared (IR) absorption spectra can be used to obtain the detailed information
about the low-energy electronic excitations in FGs. If one has the knowledge
about the influence of different stacking sequences and numbers on the IR
absorption spectra, one can identify them accurately in terms of the
experimental IR spectra. This knowledge can be obtained from theoretical
calculations, especially from first-principles calculations which have the
advantage of being empirical parameter free and therefore having the predictive
power. 
 
Motivated by the fact that natural graphite adopts the AB-stacking (Bernal)
sequence \cite{21} which gives the lowest total energy, many theoretical and
experimental studies in the past few years were focusing on the electronic and
optical properties of AB-stacking FGs \cite{22,23,24,25,26,27}, especially their infrared absorption
spectra \cite{28,29,30}. For example, it was found that the interlayer coupling in an
AB-stacking bilayer graphene (BG) leads to remarkable changes in its low-energy
dispersions as well as its IR spectra compared with those of SG
\cite{12,31,32,33,34}. On the
other hand, very recently, ABC-stacking (rhombohedral) FGs were also found
experimentally. For example, Norimatsu {\it et al} observed selective formation of
ABC-stacking graphene layers \cite{35}, and more recently, Mak {\it et al} reported
unambiguous experimental evidence for the existence of stable tetra-layer
graphenes in both AB- and ABC-stacking sequences \cite{36}. The existence of these
stable polytypes of FG provides a new possible way to tailor the electronic
structure of FGs materials for both fundamental studies and application
interests \cite{37,38,39,40}. However, because of the lack of practical samples and related
experimental observation in the past, theoretical studies on the band structure
and optical properties, especially the IR absorption spectra, of the
ABC-stacking sequence are still very limited \cite{41,42,43,44,45}, except for some simplified
tight-binding (TB) modeling \cite{46,47}. 
 
In this paper, we study the electronic structures and IR absorption spectra of
ABC-stacking tri- and tetra-layer graphenes (for simplicity, we call them ABC-3
and ABC-4, respectively) using first-principles density functional theory (DFT)
calculations. The low-energy band dispersion and the formation of the
characteristic peaks in their IR absorption spectra are analyzed in comparison
with those of the AB-stacking ones (called AB-3 and AB-4, respectively), showing
significant effects from the stacking sequence. The anisotropy of the spectra
with respect to the direction of the light electric field is also investigated
and is found to be remarkable. Our theoretical results for AB-4 and ABC-4 are in
good agreement with an very recent experimental report \cite{36}. Compared to TB model
calculations, the present calculation has the advantage of better
transferability and predictive power since it is free of any empirical
parameter. We show that, together with reliable experimental data, the
theoretical calculation of IR spectra can provide a useful tool for identifying
the stacking sequence and even the stacking number of samples. The rest part of
our paper is arranged as follows: In the next section (Section II) we give
briefly the theory and computational details; We discuss the results in Section
III, followed by a summary in Section IV .

\section{Theory and computational details}

\subsection{IR optical absorption}

The ground-state electronic structure of a FG can be calculated using density
functional theory. After the band structure $E(\mathbf{k},j)$  and corresponding eigenfunctions
are obtained, its IR absorption properties can be studied by looking at the
imaginary part of the frequency-dependent dielectric function \cite{48}: 

\begin{eqnarray}
\epsilon_2^{\alpha \beta}(\omega) &=& \frac{(2\pi e)^2}{\Omega} \lim_{q\rightarrow 0}
\frac{1}{q^2} \sum_{j_1,j_2,\mathbf{k}}2w_{\mathbf{k}} \nonumber \\
&\times&\left<u_{j_1,\mathbf{k}+\mathbf{e}_{\alpha}q} \mid u_{j_2,\mathbf{k}}\right> 
\left<u_{j_1,\mathbf{k}+\mathbf{e}_{\beta}q} \mid u_{j_2,\mathbf{k}}\right>^{*} \nonumber \\
&\times&\delta\left[ E(\mathbf{k},j_1)-E(\mathbf{k},j_2)-\omega\right],
\end{eqnarray}

where $\omega$ and $q$ are the energy and momentum of photon,
respectively; $\Omega$ denotes the
volume of the unit cell; $e$ is the charge of electron and
$w_{\mathbf{k}}$  is the weight of
k-point $\mathbf{k}$  for the k-sampling; Indices $j_1$= 1 ,2, ..., $N$ and
$j_2$ = -1, -2, ..., $-N$ denote the
conduction and valence subbands, respectively, counted from the Fermi level;
$u_{j,\mathbf{k}}$ is the cell periodic part of the wavefunction of band $j$ at
k-point $\mathbf{k}$ . Under an
electric field parallel to the graphene plane ( $\mathbf{E}\parallel \mathbf{x}$
or $\mathbf{E} \parallel \mathbf{y}$) or perpendicular to the
graphene plane ($\mathbf{E} \parallel \mathbf{z}$ ) which is studied in our work, the electrons could be excited
from the occupied valence $\pi$ bands to the unoccupied conduction $\pi^{*}$ bands. At
zero temperature, only inter-$\pi$-band excitations can occur with the excitation
energy $\omega_{\rm ex} = E(\mathbf{k},j_1) - E(\mathbf{k},j_2)$. Because photon's momentum is almost zero, the optical selection
rules is $\Delta \mathbf{k} = 0$. 

\begin{figure*}[bth]
\includegraphics[width=12cm,clip]{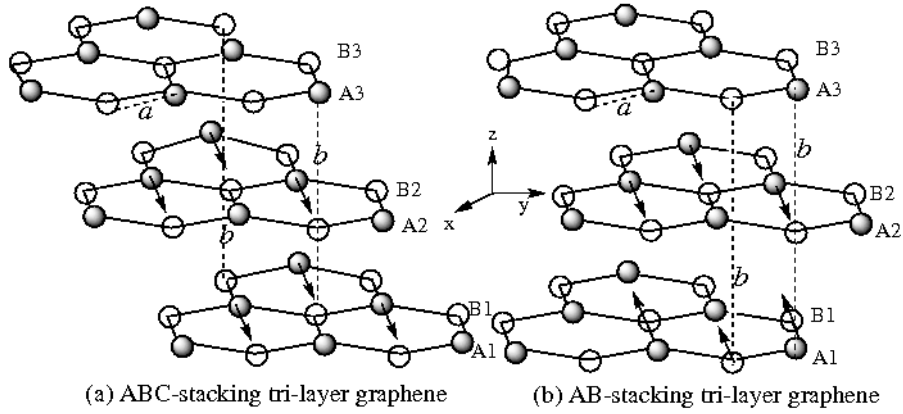}    
\caption{\label{fig:str}        
Structures of tri-layer graphenes: (a) ABC-stacking, (b) AB-stacking.
The C-C bond length is denoted by $a$ and the interlayer spacing is denoted by 
$b$.
}
\end{figure*}

\subsection{Computational details}

For band structure calculation, we use the density functional theory
implemented in the plane-wave pseudopotential formalism \cite{49} with the local
density approximation (LDA) in the Ceperley-Alder version \cite{50} for the electron
exchange and correlation. The interaction between the ions and electrons is
described by the highly accurate full-potential projected augmented wave (PAW)
method \cite{51,52} which can give a more accurate and reliable result than the
ultrasoft pseudopotential. In our calculations, the $2s$ and $2p$ orbitals of the
carbon atoms are treated as valence orbitals, and a large plane-wave cutoff of
500 eV is used throughout. A uniform grid larger than 200$\times$200$\times$1 is used in the
irreducible BZ for the k-sampling. 

A supercell geometry is constructed so that the tri- and tetra-layer graphenes
are aligned in a hexagonal supercell with the closest distance between the
adjacent FGs being at least 10{\AA} along the stacking direction ($\mathbf{z}$ direction). The
interlayer spacing is set initially to be the same as in graphite. The
structures of ABC-3 and AB-3 are shown in Figs. 1(a) and 1(b), respectively,
where there exist two C atoms, denoted by $A_i$ and $B_i$ , in a primitive cell of
each graphene layer $i$. The two polytypes can be obtained by displacing each
adjacent graphene layer continuously in one direction for the ABC stacking, and
alternatively in opposite direction for the AB stacking. We use a conjugated
gradient method to optimize the atom's positions as well as the size of the
supercell until the forces acting on all the atoms are less than 0.005 eV /{\AA}.
With the obtained equilibrium structure, we then use a more dense set of k-points 
to calculate the electronic structure and the moment matrix elements.
Finally we use the Eq.~(1) to study the IR absorption properties. 

\begin{figure*}[tb]
\includegraphics[width=14cm,clip]{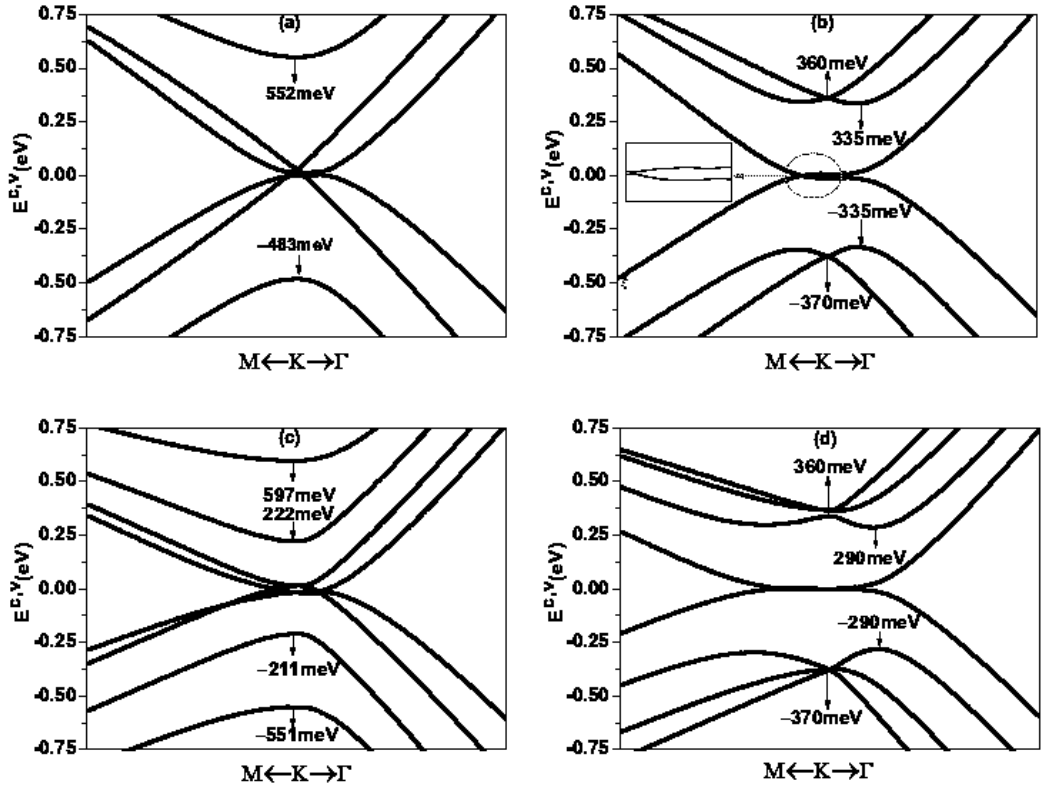}    
\caption{\label{fig:band}        
The low-energy band dispersions of the four systems studied: (a) AB-3,
(b) ABC-3, (c) AB-4, and (d) ABC-4, as used in the text.
}
\end{figure*}

\section{Results and discussion}

\subsection{Atomic structure and energy band structure}

The equilibrium C-C bond length a given by the DFT-LDA structural optimization
calculation is 1.41{\AA} for both the ABC- and AB-stacking FGs, which is almost the
same as that in bulk graphite. However, the interlayer spacing b is slightly
different for the different stacking sequences and stacking numbers: 3.31 and
3.32{\AA} for ABC-3 and ABC-4, respectively, and 3.32 and 3.34{\AA} for AB-3 and AB-4,
respectively. Our result of 3.31{\AA} for ABC-3 is in good agreement with an
available experimental value 3.335{\AA} \cite{45}, both are a bit smaller than the value
3.37{\AA} in ABC-stacking graphite \cite{53}. 

The low-energy band dispersions of the four systems are given in Figs. 2 (a) -
(d). Let's first look at the AB-stacking case (Figs. 2(a) and 2(c)). As is
evident, the low-energy band structure consists of separate single-layer-like
and bilayer-like bands \cite{44} but with some degree of distortion due to the
additional interlayer couplings. Meanwhile, the interlayer couplings reduce the
symmetry between the valence and conduction bands with respect to the Fermi
level, and also cause a small overlap between them near $E_{\rm F}$.  For example, for
AB-3, our calculation predicts different energy splittings of -483 and +552 meV
at the $K$ point for the low-lying and up-lying bands, respectively. For AB-4, the
above band splittings at the $K$ point given by our calculation are -551meV,
-211meV, 222meV, 597meV for the four bands (see Fig. 2(c)), respectively, while
they are $\pm$222 and $\pm$596 meV from a TB model calculation \cite{36}. We note that our
band structure is asymmetrical with respect to the Fermi level while the result
of the TB model \cite{36} is symmetrical. Based upon the previous experimental IR
absorption spectra of doped or gated AB-stacking FGs \cite{32,54,55}, one can deduce
that the valence and conduction bands are asymmetrical with respect to $E_{\rm
F}$. Our
result for the AB stacking is consistent with the experimental data while the
result given by the simplified TB model is not, indicating that our DFT-LDA band
structure calculation is more reasonable than the TB model calculation. 

For ABC-3 and ABC-4 which is the focus of our study, one can see in Figs. 2(b)
and 2(d) that they have strikingly different 2D band structures from those of
the AB-stacking ones, which reflect the underlying difference in
crystallographic symmetry and interlayer coupling. First, unlike in the
AB-stacking case, the band dispersions of the ABC-stacking FGs don't resemble
those of single-layer and bilayer graphenes. The different interlayer
interaction causes two intersections at the $K$ point with the energy of +360meV
and -370meV, respectively, which are absent in AB-3 and AB-4. Again, here the
band structure is slightly asymmetrical with respect to the Fermi level while it
is symmetrical in some TB model calculations ($\pm$360meV) \cite{36,37}. Second, the lower
crystallographic symmetry of the ABC-stacking shifts the extrema of the
low-lying and up-lying energy bands away from the $K$ point. In ABC-3 case, the
four shifted extrema of these "wizard-hat" bands have energies of about 335meV
away from $E_{\rm F}$, as shown in Fig, 2(b), while in ABC-4 case they are about 290meV
away from $E_{\rm F}$. Third, the interlayer interactions produce one pair of localized
flat bands near $E_{\rm F}$. For ABC-3, the two flat bands have a tiny energy gap on the
$K-\Gamma$ line while for ABC-4 they have a small overlap. Physically, these flat
bands are formed by the localized electronic states from the two outermost
graphene layers. 

\begin{figure*}[tbh]
\includegraphics[width=14cm,clip]{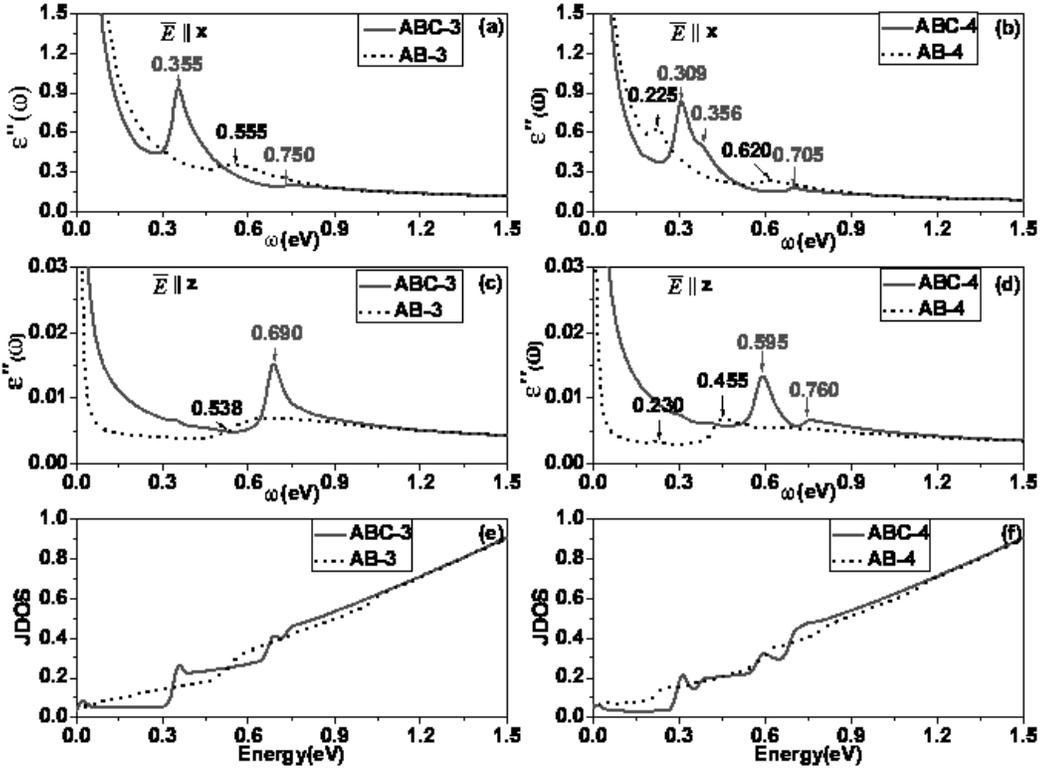}
\caption{\label{fig:spectra}
Calculated IR absorption spectra of (a) ABC-3 and AB-3 under $\mathbf{E}
\parallel \mathbf{x}$, (b)
ABC-4 and AB-4 under $\mathbf{E} \parallel \mathbf{x}$, (c) ABC-3 and AB-3 under
$\mathbf{E} \parallel \mathbf{z}$, and (d) ABC-4 and AB-4
under $\mathbf{E} \parallel \mathbf{z}$. The solid line is for ABC stacking and the dotted line is for AB
stacking. The corresponding joint densities of states (JDOS) are shown in (e)
and (f).
}
\end{figure*}

\begin{figure*}[tbh]
\includegraphics[width=12cm,clip]{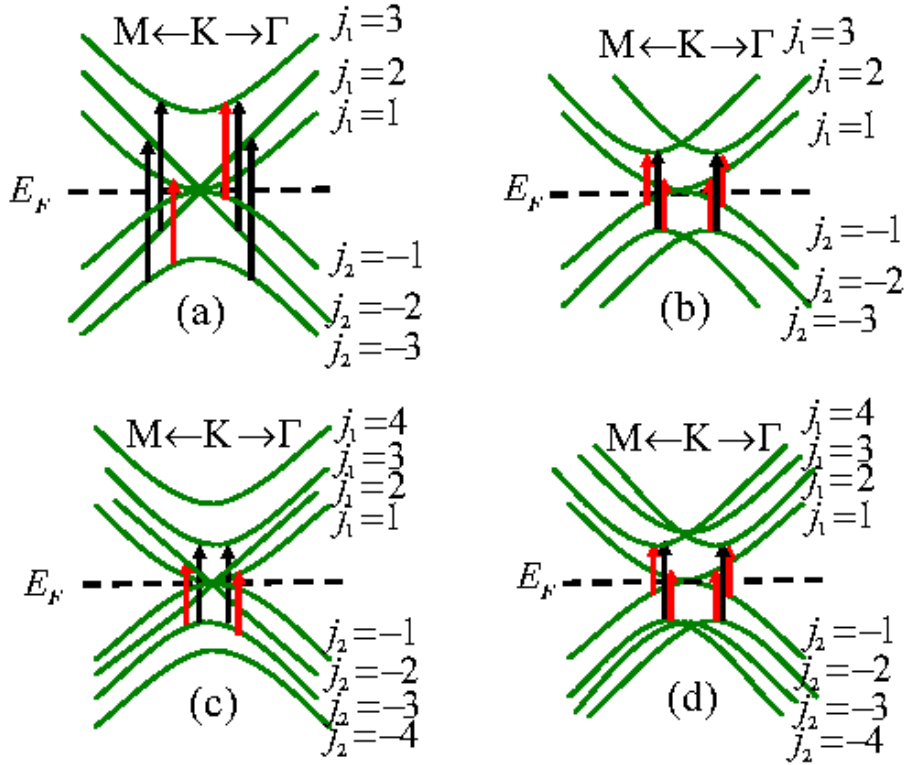}
\caption{\label{fig:transition}
Schematic drawing of the band structures of (a) AB-3, (b) ABC-3, (c)
AB-4, and (d) ABC-4, which are in the same order as in Fig. 2. Red and black
arrows denote different permitted optical transitions. Conduction bands are
denoted by index $j_1$ and valence bands are denoted by index $j_2$. 
}
\end{figure*}

\subsection{IR absorption spectra}

We calculate the IR absorption spectra of the AB- and ABC-stacking tri- and
tetra-layer FGs for two different directions of the light electric field
($\mathbf{E} \parallel \mathbf{x}$ and $\mathbf{E} \parallel \mathbf{z}$) 
and plot the results in Fig. 3 (a) - (d) together with the
corresponding joint densities of states (JDOS) in Fig. 3(e) and (f). Let us
start with the discussion about the different behavior between the ABC stacking
(the solid lines in Fig. 3) and the AB stacking (the dotted lines). We first
consider the case of $\mathbf{E} \parallel \mathbf{x}$. As one can see in Fig. 3(a), for AB-3 there exists
one characteristic peak at 555 meV, which is mainly caused by the transitions
between the valence bands $j_2$ = -3 (-1) and conduction bands $j_1$ =1 (3) with their
average energy difference being about 555 meV around the $K$ point, denoted by the
red arrows in Fig. 4(a). For ABC-3, on the other hand, its characteristic peak
is located at 355 meV. This peak is due to the transitions between valence band
$j_2$ = -2 (around the extremum) and conduction band $j_1$ =1 as well as $j_2$ =
-1 and $j_1$ = 2 (around the extremum) with their average energy difference being 355meV, as
denoted by the red arrows in Fig. 4(b). The corresponding peaks in the JDOS can
be found in Fig. 3(e). For ABC-3 the peak at 355 meV in the JDOS is due to
1D-like van Hove singularity (vHS) which was analyzed earlier by a TB model 56
and ascribed to the fact that the "wizard-hat" bands have their extrema away
from the $K$ point. In addition, for ABC-3, there also exists another weak peak at
750 meV, which is mainly caused by transitions between the two intersections
around the $K$ point, i.e., between the valence bands $j_2$ = -2, -3 and conduction
bands $j_1$ = 2, 3. Obviously, the characteristic peak at 355meV of ABC-3 has a
red-shift of 200meV compared with that at 555 meV of AB-3 while the peak at
750meV of ABC-3 is absent in AB-3. For AB-4, as one can see in Fig. 3(b), there
are two characteristic peaks, one at 225 meV and the other at 620 meV. The peak
at 225 meV is mainly caused by the transitions between the valence bands $j_2$ = -3
(-3) and conduction bands $j_1$ =1 (2) with the interlayer-coupling induced
splitting energy between the pairs being about 225 meV, which are denoted by the
red arrows in Fig. 4(c). The peak at 620 meV is due to the transitions between
the valence bands $j_2$ = -2, -1, -4 and conductance bands $j_1$ = 4, 4, 1, respectively.
For ABC-4, on the other hand, there is one major peak at 309 meV which is mainly
caused by the transitions between the valence band $j_2$ = -2 (around the extremum)
and conduction band $j_1$ = 1 as well as $j_2$ = -1 and $j_1$ = 2 (around the extremum) with
the average energy difference 309 meV, denoted by the red arrows in Fig. 4(d).
The corresponding peak in the JDOS (1D-like vHS) can be observed at around 309
meV in Fig. 3(f). Besides the major characteristic peak there also exist two
weaker structures in the IR spectra of ABC-4. One is a "shoulder structure" at
356 meV, which is mainly caused by the transitions between the valence bands
$j_2$ = -3 (-1) and conduction bands $j_1$ = 1 (3) around the $K$ points. The other is a
weaker peak at 705 meV which is induced by the transition between the
intersections of the valence bands $j_2$ = -2 (-3) and conduction bands $j_1$ = 3 (2)
around the $K$ points. Overall, the three structures in the IR spectrum of ABC-4
are blue-shifted with respect to the two structures of AB-4. Specifically, the
major peak at 309 meV has a blue-shift of 84 meV with respect to the peak at 225
meV of AB-4. The weak peak at 705 meV of ABC-4 has a blue-shift of 85 meV with
respect to the peak at 620 meV of AB-4. While the "shoulder structure" at 356
meV is absent in AB-4.  

Next, we consider the case of $\mathbf{E} \parallel \mathbf{z}$ (see Figs. 3(c) and (d)). For AB-3, a jump
lies at 538 meV, which is caused by transitions between valence bands $j_2$ = -2 (-3)
and conduction bands $j_1$ = 3 (2) near the $K$ point with their average energy
difference being about 538 meV, denoted by the black arrows in Fig. 4(a). The
corresponding jump structure in the JDOS can be seen in Fig. 3(e). In contrast,
for ABC-3, a characteristic peak appears at 690 meV. The transitions behind are
between the valence band $j_2$ = -2 (around the extremum) and conduction band
$j_1$ = 2 (around the extremum) as well as $j_2$ = -3 and $j_1$ = 3, denoted by the black arrows in
Fig. 4(b). This peak corresponds to the 1D-like vHS at 690 meV in the JDOS (see
Fig. 3(e)), which is quite different from the 2D systems' characteristic jump
structure (or call step singularity) at 538 meV of AB-3.  For AB-4, as can be
seen in Fig. 3 (d), there is a weak peak at 230 meV and a major peak at 455 meV
which is induced by the transition between $j_2$ = -3 and $j_1$ = 3, denoted by the black
arrows in Fig. 4(c). In Fig. 3(f) one can see the corresponding features in the
JDOS for these two peaks. For ABC-4, on the other hand, a major peak is at 595
meV, as shown in Fig. 3(d), which is due to the transitions between the extrema
of valence band $j_2$ = -2 and conduction band $j_1$ = 2, denoted by the black arrows in
Fig. 4(d). The corresponding peak in the JDOS (1D-like vHS) is shown in Fig.
3(f). Another weaker peak at 760meV (Fig. 3(d)) is caused by the transition
between $j_2$ = -3 and $j_1$ = 3 around the $K$ point. Overall, the peaks in the IR spectrum
of ABC-4 are blue-shifted with respect to those of AB-4. 

The different IR absorption behavior between the ABC- and AB-stacking FGs can
be understood by considering the difference in their electronic structure. As
mentioned previously, the ABC stacking has a lower crystallographic symmetry
compared with the AB stacking. As a result, the extrema of its low- and up-lying
energy bands are shifted away from the $K$ point, which gives rise to the 1D-like
vHS in their JDOS and prominent optical absorption peaks. These prominent peaks
are in strong contrast with the much weaker intensity structures in the IR
spectra of the AB stacking, and also located at very different frequencies . 

So far, almost all experiments \cite{32,36,54,55} and theoretical TB model
calculations \cite{15,46,47} considered only the case of $\mathbf{E} \parallel
\mathbf{x}$ in investigating the IR
spectra of FGs, while the case of $\mathbf{E} \parallel \mathbf{z}$ was little
concerned \cite{34} although the setup
of  $\mathbf{E} \parallel \mathbf{z}$ is feasible in experiment and can provide more information, as already
implied in our previous discussion. Below we discuss in more details the effect
of different directions of the light electric filed on the IR absorption
spectra. It is evident from Fig. 3 that the IR absorption spectra show a strong
anisotropy between  $\mathbf{E} \parallel \mathbf{x}$ and $\mathbf{E} \parallel
\mathbf{z}$. For ABC-3, the position of its major
characteristic peak (at 690 meV) under $\mathbf{E} \parallel \mathbf{z}$ has a large blue-shift of 335 meV
compared with the case of $\mathbf{E} \parallel \mathbf{x}$. In addition, there is a weak peak at 750 meV
under $\mathbf{E} \parallel \mathbf{x}$, which is absent under $\mathbf{E}
\parallel \mathbf{z}$. For AB-3 the peak at 555 meV under $\mathbf{E} \parallel
\mathbf{x}$ is replaced by a jump, instead of a peak, around 538 meV, as shown in Fig. 3(c).
Similar to the tri-layer systems, the tetra-layer systems also show a strong
anisotropy effect between $\mathbf{E} \parallel \mathbf{x}$ and $\mathbf{E}
\parallel \mathbf{z}$, particularly ABC-4: Its characteristic
peaks are significantly blue-shifted from $\mathbf{E} \parallel \mathbf{x}$ to
$\mathbf{E} \parallel \mathbf{z}$, by up to 286 meV. One
thing to note is that, compared with the AB stacking, the ABC stacking has a
much stronger anisotropy effect causing large blue-shifts for its major peaks
from $\mathbf{E} \parallel \mathbf{x}$ to $\mathbf{E} \parallel \mathbf{z}$ 
(see Fig. 3 (a) vs. (c), and (b) vs. (d)). Physically, this
can be understood also by considering its lower crystallographic symmetry which
results in a very different band structure including the "wizard-hat" bands.
This band structure provides two sets of transitions with significantly
different transition energies for $\mathbf{E} \parallel \mathbf{x}$ and
$\mathbf{E} \parallel \mathbf{z}$, respectively (see Fig. 4, red
arrows vs. black arrows). 

Finally, we would like to discuss how the number of stacking layers affects the
IR spectra. As can be seen in Fig. 3, the thickness effect ((a) vs. (b) and (c)
vs. (d)) is not so significant compared to the effect from stacking sequence.
For the ABC stacking, all the characteristic peaks have a red-shift as the
stacking number increases. For the AB stacking, however, the trend is not so
clear. Under $\mathbf{E} \parallel \mathbf{x}$ the maximum shift in peak position is about 45meV between
ABC-3 and ABC-4. This shift is enhanced to more than 90meV under $\mathbf{E}
\parallel \mathbf{z}$.
Physically, this enhanced effect can be understood by considering the
interaction between the light electric field and the graphene sheets. In the
former case the field will induce an intra-layer polarization which is less
sensitive to the thickness while in the latter the field will induce an
inter-layer polarization which is more sensitive to the thickness. 

\begin{figure}[t]
\includegraphics[width=8.0cm,clip]{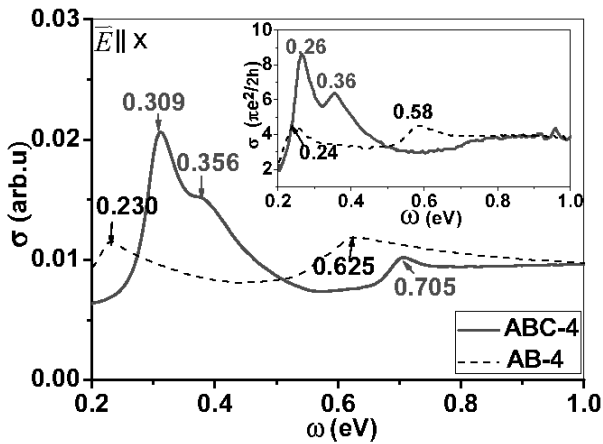}
\caption{\label{fig:conductance}
IR optical conductivity spectra of ABC-4 (solid line) and AB-4 (dashed 
line) under $\mathbf{E} \parallel \mathbf{x}$. The inset shows the corresponding experimental results
reported by Mak {\it et al} in Ref.[36]. The positions of the peaks are denoted.
}
\end{figure}

From our discussion, one can see that different stacking sequences of FGs have
very significant effects on their IR absorption spectra, leading to peaks at
different frequencies and with different intensities. Furthermore, our
calculation shows that the IR spectra are also sensitive to the direction of the
light electric field, especially for the ABC stacking sequence. In the case of
$\mathbf{E} \parallel \mathbf{z}$ the thickness of an ABC-stacking sample also plays an important role. The
high sensitivities of the IR absorption spectra to the different factors provide
important information for identifying the stacking sequence and stacking number
of an experimental sample by comparing the experimental spectra with the
calculated ones. Below, we compare our theoretical result with an experimental
report \cite{36} which is, to the best of our knowledge, the only available one in
literature for ABC-stacking FGs so far. In order to have a reasonable
comparison, we plot the IR optical conductivity spectra $\sigma(\omega)=\omega
\epsilon_2 (\omega)/4\pi$ \cite{57} in Fig. 5 for ABC-4
and AB-4 under $\mathbf{E} \parallel \mathbf{x}$. One can see that our theoretical results are in good
overall agreement with the experiment data (see the inset in Fig. 5) for both
ABC-4 and AB-4. The positions of the peaks and their relative intensities as
well as the overall shape of the two experimental spectra are all well
reproduced. Quantitatively, however, there are still some discrepancies between
theory and experiment. For AB-4 the maximum difference in the peak position is
about 45 meV while for ABC-4 it is up to 50 meV. Additionally, for ABC-4 our
calculation predicts a weak peak at 705 meV, which was also found in the
simplified TB's model calculation \cite{36} but with a slight shift (670 meV). However,
this peak was not observed in the experiment. How to understand these
quantitative discrepancies is still an open problem. It may be due to some
experimental environment which has not been taken into account in our DFT
calculation, such as doping, disorder, defect effects. The experimentally missed
peak at 705 meV, on the other hand, may be due to temperature broadening. To
clarify the quantitative inconsistency between theory and experiment, further
theoretical work and more experimental measurements are desirable. 

\section{Summary}

We have studied the IR absorption spectra of ABC-stacking tri- and tetra-layer
FGs using a first-principle method for two different directions of the light
electric field ($\mathbf{E} \parallel \mathbf{x}$ and $\mathbf{E} \parallel
\mathbf{z}$), and compare them with those of AB-stacking ones.
The findings are as follows. 1) The ABC-stacking sequence causes great different
band structures, inducing characteristic peaks at different positions and with
different intensities, no matter $\mathbf{E} \parallel \mathbf{x}$ or
$\mathbf{E} \parallel \mathbf{z}$. 2) The IR spectra show a strong
anisotropy effect between $\mathbf{E} \parallel \mathbf{x}$ and $\mathbf{E}
\parallel \mathbf{z}$. 3) The IR spectra are much more
sensitive to the stacking number under $\mathbf{E} \parallel \mathbf{z}$ because of the induced inter-layer
polarization. 4) Our calculated IR optical conductivity spectra for the
tetra-layer graphenes under $\mathbf{E} \parallel \mathbf{x}$ are well consistent with a recent experimental
observation. The significant effects of the different factors on the IR spectra
of FGs provide useful tools for identifying their stacking sequence and stacking
number. 


This work was supported by Shanghai Pujiang Program under Grant 10PJ1410000 and
the MOST 973 Project 2011CB922200. 


\end{document}